\documentclass[twocolumn]{aastex63}

\usepackage{lineno}

\usepackage{amsmath}

\usepackage{xspace}

\usepackage{color}

\newcommand{\MESA}{{\texttt{MESA}}\xspace{}}

\newcommand{\mergprob}{{merger fraction}}
\newcommand{\mergprobeq}{\ensuremath{F(\mathrm{merge})}}
\newcommand{\highlight}{ }

\newcommand{\be}{\begin{enumerate}}
\newcommand{\ee}{\end{enumerate}}

\shorttitle{}
\shortauthors{Gallegos-Garcia et al.}

\begin{document}

\title{Evolutionary Origins of Binary Neutron Star Mergers: Effects of Common Envelope Efficiency and Metallicity} 
\author[0000-0003-0648-2402]{Monica Gallegos-Garcia}
\affiliation{Department of Physics and Astronomy, Northwestern University, 2145 Sheridan Road, Evanston, IL 60208, USA}
\affiliation{Center for Interdisciplinary Exploration and Research in Astrophysics (CIERA),1800 Sherman, Evanston, IL 60201, USA}

\author[0000-0003-3870-7215]{Christopher P L Berry}
\affiliation{Department of Physics and Astronomy, Northwestern University, 2145 Sheridan Road, Evanston, IL 60208, USA}
\affiliation{Center for Interdisciplinary Exploration and Research in Astrophysics (CIERA),1800 Sherman, Evanston, IL 60201, USA}
\affiliation{SUPA, School of Physics and Astronomy, University of Glasgow, Glasgow G12 8QQ, UK}

\author[0000-0001-9236-5469]{Vicky Kalogera}
\affiliation{Department of Physics and Astronomy, Northwestern University, 2145 Sheridan Road, Evanston, IL 60208, USA}
\affiliation{Center for Interdisciplinary Exploration and Research in Astrophysics (CIERA),1800 Sherman, Evanston, IL 60201, USA}

\begin{abstract}
The formation histories of compact binary mergers, especially stellar-mass binary black hole mergers, have recently come under increased scrutiny and revision.
We revisit the question of the dominant formation channel and efficiency of forming binary neutron star mergers.
We use the stellar and binary evolution code \texttt{MESA}, and implement a detailed method for common envelope and mass transfer. 
We preform simulations for donor masses between $7$--$20 M_{\odot}$ with a neutron star companion of $1.4M_{\odot}$ and $2.0M_{\odot}$, at two metallicities, using varying common envelope efficiencies, and two different prescriptions to determine if the donor undergoes core collapse or electron capture given their helium and carbon--oxygen cores.
In contrast to the case of binary black hole mergers, for a neutron star companion of $1.4M_{\odot}$, all binary neutron star mergers are formed following a common envelope phase. 
For a neutron star mass of $2.0M_{\odot}$ we identify a small subset of mergers following only stable mass transfer  if the neutron star receives a natal kick sampled from a Maxwellian distribution with velocity dispersion $\sigma = 265 \ \rm{km \ s}^{-1}$.
Regardless of supernova prescription, we find more binary neutron star mergers at subsolar metallicity compared to solar.

\end{abstract}

\keywords{Gravitational wave sources (677); Neutron stars (1108); Stellar evolutionary models (2046); Common envelope evolution (2154); Roche lobe overflow (2155)}

\newpage

\section{Introduction}

Binary systems hosting neutron stars (NSs) fall in the intersection of many observed phenomena. 
For several decades, observations of radio and X-ray pulsars \citep{Joss1984,Lorimer2008}, as well as X-ray binaries \citep{Verbunt1993} have offered many insights into the formation of NSs in binaries. 
More recently, the LIGO Scientific, Virgo and KAGRA Collaboration have detected gravitational waves (GWs) from coalescing binary NSs (BNSs) and neutron star--black hole (NSBH) binaries \citep[][]{Abbott2017_GW170817,Abbott2021_2nd_part_GWTC3,Abbott2021_2NSBH,Abbott2021_first_half_GWTC3_p2}. 
The upcoming fourth observing run is expected to further increase the GW sample \citep{Abbott2018_prospects_aLIGO,Abbott2021_pop_through_GWTC3}.
Unlike typical merging binary black holes (BBHs), BNS mergers can also be observed via electromagnetic radiation providing a wealth of multimessenger data \citep[][and references therein]{Metzger2017_kilonova_review,Abbott2017_Multimessenger_BNS,Abbott2017_GW_and_GRs,Margutti2021}.
Given the impact that BNS mergers have across astronomy, the question of how they form remains of importance.

Merging BNSs are believed to primarily form through isolated binary evolution \citep{Postnov2014,Tauris2017}, and only rarely form dynamically, if at all \citep{Ye2020}.
In the isolated evolution scenario, two massive main sequence (MS) stars are initially detached in an orbit. 
Throughout their life, radial expansion and a close orbit can lead to episodes of stable and unstable mass transfer (MT).
A common envelope (CE) phase, which is initiated by unstable MT, occurs when the binary becomes embedded within a shared stellar envelope \citep{Paczynski1976_CE,vandenHeuvel1976_CE}. 
In this phase orbital energy is used to unbind the envelope \citep[][]{Webbink1984}.
Following a CE, the exposed stellar core will explode in a supernova (SN) leaving behind a NS. 
If the explosion is asymmetric, the NS receives a kick \citep{Janka1994,Burrows1996,Janka2013}, imparting eccentricity onto the orbit and potentially disrupting the binary.
These phases of binary evolution are simulated with both rapid population synthesis and detailed modeling, which have helped constrain key uncertainties such as the formation rates of BNSs and NSBHs \citep[e.g.,][]{MandelBroekgaarden2022}. 
By comparing to observations we may learn more about key phases and properties in the isolated formation channel such as kicks, CE, and the NS mass distribution \citep[e.g.,][]{Wong2010,Tauris2017}.
The CE phase in particular plays a crucial role in the isolated evolution scenario as it is believed to be the primary way to form BNS consistent with observed short-period systems and BNS close enough to merge within a Hubble time.
Unfortunately critical phases of CE remain poorly understood \citep{Ivanova2013,IvanovaRicker2020}.

In the case of BBH mergers, both theoretical and observations results indicate that they can form from more than one channel \citep{Abbott2021_pop_through_GWTC3,Zevin2021_one_channel,Mapelli2021,Mandel2022}.
Until recently, the formation of BBHs through the isolated formation channel was thought to be dominated by CE evolution. 
Studies have since shown that that stable MT can play a larger or dominant role in the formation of BBH and BBH mergers and in some cases have emphasised that the CE survival in BBH merger progenitors may be over estimated in rapid population synthesis codes \citep{vandenHeuvel2017,Pavlovskii2017,neijssel_effect_2019,Klencki2021,olejak2021impact,Marchant2021,Gallegos-Garcia2021}.
A natural question is whether the same is true for the formation of BNS mergers.
In other words, what is the dominant formation channel for BNS mergers and how efficient is CE evolution at forming BNS mergers?

While we may not yet fully understand the intricacies of the CE phase, such as its efficiency at unbinding the envelope, we can use detailed stellar and binary evolution models to minimize uncertainties related to incomplete modeling and to simulate how the stellar envelope responds to MT.
In this paper, we examine the formation of BNS mergers for the isolated formation scenario after the first NS is formed.  
Following our previous study of BBH evolution \citep{Gallegos-Garcia2021}, we use \MESA{} simulations with the detailed method of \citet{Marchant2021} to model CE and MT in our simulations.
In Sec.~\ref{sec:methods} we briefly describe our main choices and modification for stellar and binary physics models, in Sec.~\ref{sec:MESA_results} we describe our main results, and provide a summary in Sec.~\ref{sec:conclusions}.

\section{Method}\label{sec:methods}

For our simulations we use \MESA{} version 12115 and employ the detailed MT and CE method of \cite{Marchant2021}. 
We primarily follow the stellar and binary physics prescriptions and assumption used in \cite{Gallegos-Garcia2021}. 
In Sec.~\ref{subsection:binary_physics} we briefly discuss our main choices and modifications for stellar and binary physics models, and summarize how we quantify BNS mergers, taking into account a distribution of natal kicks. In Sec.~\ref{section:model_variations} we briefly describe our model variations, which are chosen to explore uncertainties in binary evolution such as CE efficiency and metallicity. 

\subsection{Stellar and Binary Evolution Physics} \label{subsection:binary_physics}

We initialize our standard model at a metallicity of $Z = Z_{\odot}$, defining $Z_{\odot} = 0.0142$ and $Y_{\odot} = 0.2703$ \citep{2009Asplund}. 
We specify the helium fraction as $Y = Y_\mathrm{Big\ Bang} +  ( Y_{\odot} - Y_\mathrm{Big\ Bang} ) Z/Z_{\odot}$, where $Y_\mathrm{Big\ Bang} = 0.249$ \citep{Planck2016}. 
We use the \texttt{Dutch} stellar winds prescription in \MESA{}, which is based on \citet{Glebbeek2009}. 
This prescription incorporates \citet{Vink2001_hot_highH} for effective temperatures of $T_{\rm eff} > 10^4~\mathrm{K}$ and surface hydrogen mass fraction of $H>0.4$; \citet{Nugis2000_hot_lowH} for $T_{\rm eff} > 10^4~\mathrm{K}$ and $H<0.4$ (Wolf--Rayet stars), and \citet{deJager1988} for $T_{\rm eff} < 10^4~\mathrm{K}$. 
We evolve models until they reach core carbon depletion (central $^{12}$C abundance  $< 10^{-2}$).
NS accretors, which we treat as point masses in \MESA{}, begin with a mass of $1.4M_{\odot}$ in our standard model. 
We assume the orbit of the binary remains circular but allow eccentricity to be imparted onto the binary at the time of a SN explosion.

For our treatment of MT and CE, we implement the detailed method of \cite{Marchant2021}. 
The MT is an extension of \citet{Eggleton1983} and \citet{KolbRitter1990}, and accounts for potential outflows from outer Lagrangian points with the Roche lobe radius approximated using \citet{Eggleton1989}.
The CE method follows the standard energy prescription \citep{Webbink1984,deKool1990,DewiTauris2000} but solves for the binding energy of the donor using the stellar profile and  self-consistently determines the core-envelope boundary. 
We assume MT is unstable when the donor exceeds a MT rate of $\dot{M} > 1 M_{\odot}~\mathrm{yr}^{-1}$. 
At the onset of CE, the binding energy is calculated using 
\begin{equation}
    E_\mathrm{bind} (m) =  \int^{M_{\rm donor}}_{m} \left( - \frac{G m'}{r} + \alpha_{\rm th} u\right) {\rm d}m',
\end{equation}
where $m$ is an arbitrary mass coordinate, $u$ is the specific internal energy of the gas, and $\alpha_\mathrm{th}$ is the fraction of $u$ that can be used for the ejection of the envelope. 
During CE, mass is stripped off the donor at a high rate. The prescription uses $E_\mathrm{bind} (m)$, along with the CE efficiencies and orbital parameters, to iteratively update the orbital separation until the system detaches.
In preliminary tests, we found that the addition of the specific internal energy $\alpha_\mathrm{th}$ to unbind the core increased the survivability of CE for the low-mass donors by almost $50\%$.
As a result, we use $\alpha_{\mathrm {th}} = 1$ in our standard model to include these effects into our simulations. 

For Eddington-limited accretion, the rate of accretion onto a compact object depends on the dimensionless constant $\eta = G M_{\mathrm{acc}}/R_{\mathrm{acc} }c^2$, where $M_{\mathrm{acc}}$ and $R_{\mathrm{acc}}$ are the mass and radius of the accretor. 
This value sets how efficiently accreted mass can be converted into radiation. 
We use a constant $R_{\mathrm{acc}} = 12.5~\mathrm{km}$ following \citet{POSYDON2022}.

For SN explosions, which determine both the final masses and natal kicks of the resulting NS, we implement two different thresholds to determine if the donor undergoes core-collapse (CC) or electron-capture SN (ECSN).
In one case, we assume the donor star collapses in an ECSN if at the end of the simulation the He-core mass is between $1.4 M_{\odot} < M_{\mathrm{He,core}}  <2.5 M_{\odot}$ \citep{Podsiadlowski2005}.
In the second case, we limit ECSNe to donor stars with carbon--oxygen cores between $1.37 M_{\odot} < M_{\mathrm{CO,core}} < 1.43 M_{\odot}$ \citep{Tauris2015}.
We assume SN explosions will form a $1.4 M_{\odot}$ NS.
Donors below these ranges are assumed to form white dwarfs (WDs).

For SN natal kicks of the NS, we draw kick velocity magnitudes from a Maxwellian distribution with a velocity dispersion determined by the type of SN explosion.
For donors that collapse in EC we use $\sigma_{\mathrm{ECSN}} = 20~\mathrm{km\,s}^{-1}$ following \citet{GiacobboMap2019_ecsn}, and use $\sigma_{\mathrm{CCSN}} = 265~\mathrm{km\,s}^{-1}$ for CC following \cite{Hobbs2005}. 
For each newly-formed NS we sample $2000$ kick velocity magnitudes from this distribution, which we use to calculate the post-SN orbit following \citet{1996Kalogera}.
For each BNS system in our grid, we use the distribution of 2000 post-SN orbits to calculate a distribution merger times due to the radiation of GWs \citep{Peters1964}. 
Our results are given in terms of a \mergprob{} for each grid point \mergprobeq{}, which is the fraction of kick samples that led to BNS mergers within a Hubble time.

We simulate donors with masses at $7 M_{\odot}$, $8 M_{\odot}$, $10 M_{\odot}$, $12.5 M_{\odot}$, $15 M_{\odot}$ and $20 M_{\odot}$.
The selection of donor masses was determined from a preliminary study using three SN prescriptions on single stars: \citet{Fryer2012}-rapid, \citet{Fryer2012}-delayed and \citet{Sukhbold2016}.
For the stellar models used, only \cite{Fryer2012}-rapid produced NSs from ZAMS progenitors between roughly $15$--$20 M_{\odot}$. 
As a result, we concentrate on the regions where multiple models predict NSs.

\subsection{Model Variations} \label{section:model_variations}

For our standard model we use solar metallicity $Z~=~Z_{\odot}$, a CE efficiency of $\alpha_{\mathrm{CE}}~=~1$, and a NS mass of $1.4M_{\odot}$. 
In addition to this model, we consider models at solar metallicity with $\alpha_\mathrm{CE} = 2$ and $\alpha_\mathrm{CE} = 0.5$, and a model at subsolar metallicity $Z=0.1 Z_{\odot}$ with $\alpha_{\mathrm{CE}} = 1$.
For these four models we apply two prescriptions for determining which explosions result in an ECSN as described above.
To consider the affects of the mass ratio $q=M_{\mathrm{NS}}/M_{\mathrm{donor}}$, we show one additional model with a donor mass of $8M_{\odot}$ and a NS mass of $2.0M_{\odot}$ at $Z=0.1 Z_{\odot}$ and $\alpha_{\mathrm{CE}} = 1$.

For each model we compute a grid of simulations consisting of a MS donor with a NS accretor in a circular orbit. 
The grids span an initial period range between $-0.1<\log_{10}(P_{\mathrm{orb,i}}/\mathrm{days})<3.5$. 

\section{Results} \label{sec:MESA_results}

\begin{figure*}
\centering
\includegraphics[width=0.99
\textwidth]{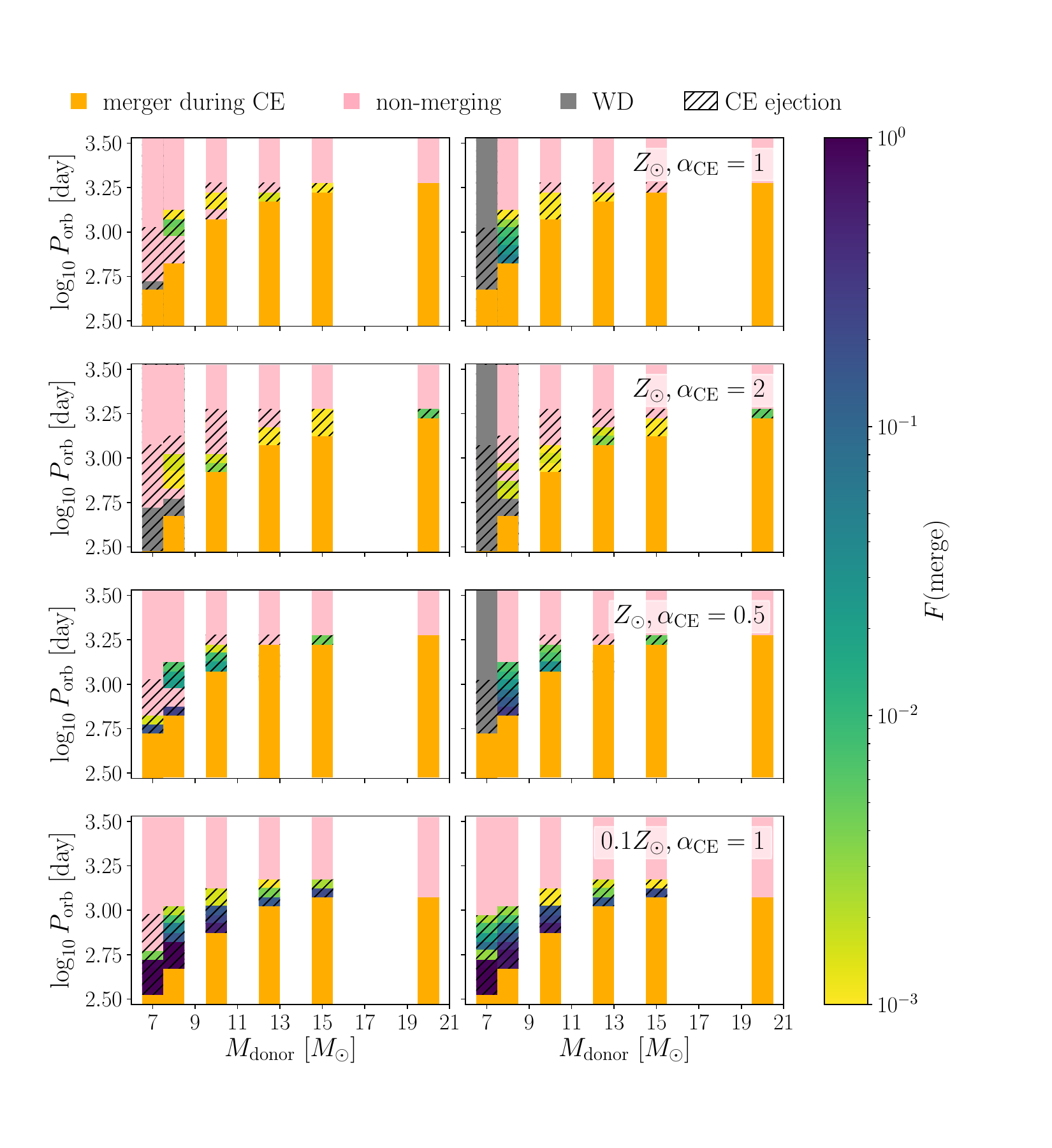}
\caption{Final outcomes of our simulations of NS--MS binaries with $M_{\mathrm{NS}}=1.4 M_{\odot}$. 
We use two different prescriptions to determine if the explosion is a CCSN or ECSN: following \citet[][\emph{left}]{Podsiadlowski2005} and following \citet[][\emph{right}]{Tauris2015}. 
\emph{First row}: models ran at solar metallicity $Z_{\odot}$ and $\alpha_{\mathrm{CE}} = 1$; \emph{second row}:  $Z_{\odot}$ and $\alpha_{\mathrm{CE}} = 2$;  \emph{third row}: $Z_{\odot}$ and $\alpha_{\mathrm{CE}} = 0.5$, and \emph{bottom row}: subsolar metalicity $0.1Z_{\odot}$ and $\alpha_{\mathrm{CE}} = 1$. 
Colors on the color scale indicate the fraction of BNS mergers. 
Regardless of SN prescription, simulations at subsolar metallicity have a higher fraction of forming BNS mergers, followed by simulations at solar metallicity with $\alpha_{\mathrm{CE}} = 0.5$.
}
\label{fig:grids}
\end{figure*}

Our focus is to identify the main evolutionary channel through which BNS mergers form in our models.
We examine our \MESA{} models for BNS formation in terms of the evolutionary class they belong to: {\it merger during CE}, {\it non-merging systems}, and {\it BNS mergers}. 
In Sec.~\ref{sec:1p4NS} and Sec.~\ref{sec:2p0NS} we describe results for simulations with a NS mass of $1.4 M_{\odot}$ and $2.0 M_{\odot}$ respectively.

\subsection{BNS simulations of binaries with $1.4M_{\odot}$ NS companions} \label{sec:1p4NS}

In Fig.~\ref{fig:grids} we show the different classes of \MESA{} simulations as a function of initial orbital period $P_{\mathrm{orb}}$ and donor mass.
The left column in Fig.~\ref{fig:grids} corresponds to simulations where ECSNe were determined following \citet{Podsiadlowski2005}.
The right column uses more narrow limits for ECSNe following \citet{Tauris2015}. 
The majority of our simulations end in forming a BNS, but at sufficiently low donor masses a WD is formed instead of a NS. 
Envelope stripping during CE can further allow for the formation of a WD by decreasing the final donor and core masses. 
Hence, in our models we find WDs tend to form at the lowest donor masses and highest $\alpha_{\mathrm{CE}}$. 
The transition between forming a WD and forming a NS is uncertain. 
Our results are estimates of this boundary, and a more exhaustive analysis is needed to accurately models this transition, especially following episodes of significant mass loss.
We do not calculate a merger fraction for systems containing a WD.

For the BNSs, we identify the three final outcome classes using the system's CE evolution outcome and \mergprob{} \mergprobeq{}. 
Regardless of final outcome, we use a hatch pattern to denote systems that successfully eject the envelope during the CE phase.
All the models follow similar overall trends.
The three classes of BNSs in Fig.~\ref{fig:grids} are:
\begin{enumerate}
    \item {\it Merger during CE} --- The (orange) region corresponding to systems that merge during CE is the largest class in all our models.
    It occupies the bottom portion of each grid at $P_{\mathrm{orb}} \lesssim 1000$ days. 
It is the largest for two reasons. 
First, the NS mass we use in this grid leads to mass ratios $q < 0.175$, which for a wide range of initial orbital periods results in a high mass transfer rate and CE. 
Second, in order to successfully eject the envelope, a deep convective envelope at the onset of CE is needed \citep{Marchant2021, Klencki2021} as well as enough initial orbital energy, which occurs at large initial orbital periods. 
As a result, CE leads to a merger for a wide range of periods.
With increasing donor mass, this outcome occurs at larger initial periods.
This is because more massive stars expand more and therefore can initiate CE at larger initial orbital periods.

\item 
{\it Non-merging systems} --- The (pink) region corresponding to non-merging systems occurs at the largest initial orbital periods. 
These non-merging systems include wide BNS binaries that do not merge within a Hubble time and binaries that are disrupted during the SN explosion.

\item 
{\it BNS mergers} --- Between the systems that merge during CE and non-merging systems, we find a narrow region in the $M_{\mathrm{donor}}$--$P_{\mathrm{orb}}$ parameter space where our simulations produce BNS mergers following CE.
These simulations correspond to binary systems where, after receiving a distribution of kick velocities, there was in at least one BNS merger within a Hubble time.
The colors of these simulations correspond to the \mergprob{} \mergprobeq{}, the fraction of kicks that lead to BNS mergers. 
This region becomes narrower with increasing donor mass. 
This because as the donor mass increases, the range in radial evolution at which the star develops a convective envelope decreases \citep[e.g.,][Figure 19]{Marchant2021}. 
Since a deep convective envelope is needed for CE ejection, and because different initial orbital periods correspond to different stellar radii at the onset of CE, we find fewer successful ejections at larger masses. 

\end{enumerate}
All BNS mergers in our models from Fig.~\ref{fig:grids} form  following CE. 
In the following we discuss how the efficiency of forming these mergers through CE is affected by the model variations that we consider.

In the top row of Fig.~\ref{fig:grids} we show the final outcomes for our standard model. 
For this model, the \emph{pre}-SN orbital periods of systems with successful CE ejections are too wide to form merging BNS within a Hubble time.
Once the SN kicks are applied, we find a maximum merger fraction of $\mergprobeq{}=0.06$, which occurs when using the ECSN prescription following \citet{Tauris2015}. 
Under this prescription, none of the NSs form via an ECSN and therefore receive larger CCSN kicks. 
These larger kicks perturb the wide pre-SN orbit with enough eccentricity for a few systems to merge within a Hubble time.
Under the SN prescription following \cite{Podsiadlowski2005}, nine out of the $19$ NSs formed following CE explode in an ECSN.
They receive low natal kicks that preserve the wide pre-SN orbit and do not merge within a Hubble time.

WDs form at low donor masses where their core masses can fall below the ECSN range. 
For the prescription of \citet{Tauris2015}, all $7 M_{\odot}$ donors that do not merge during the CE phase form WDs; however, for the prescription of \citet{Podsiadlowski2005}, only one donor at $7 M_{\odot}$ forms a WD.

Although implementing a different ECSN criterion impacts only a few systems in our grid, the affected binaries are those with the highest merger fraction, and are lower mass stars that are more abundant in the Universe. 
Our results are sensitive to the velocity magnitude of the SN kick, which are determined by which systems undergo ECSN. 
For example, if all BNS systems receive low ECSN kicks, the \mergprob{} for all systems drops to $\lesssim10^{-3}$.
This indicates that determining which systems undergo ECSN is important for tracing the formation of BNS mergers \citep{Vigna2018,GiacobboMap2019_ecsn}.

The second row of Fig.~\ref{fig:grids} corresponds to models at solar metallicity but with $\alpha_{\mathrm{CE}} = 2$.
The maximum \mergprob{} for this model is smaller compared to the first row ($\alpha_{\mathrm{CE}} = 1$), with a value of $\mergprobeq = 0.006$.
This likely occurs for two reasons.
First, for systems that survived CE at $\alpha_{\mathrm{CE}} = 1$, a larger $\alpha_{\mathrm{CE}}$ now results in wider post-CE separations which are less likely to lead to a merge within a Hubble time.  
Second, although $\sim70\%$ more systems eject the envelope compared to simulations at $\alpha_{\mathrm{CE}}=1$, they have wider post-CE separations than the binaries with highest merger fractions at $\alpha_{\mathrm{CE}}=1$. 
On the other hand, as shown in the third row, simulations with solar metallicity but at $\alpha_{\mathrm{CE}} = 0.5$, show the opposite overall effect.
At this lower efficiency, more orbital energy is required to unbind the envelope and the post-CE separations are shorter compared to $\alpha_{\mathrm{CE}} = 1$. 
As a result, we find higher merger fractions with a maximum of $\mergprobeq = 0.3$. 
Additionally, unlike simulations at $\alpha_{\mathrm{CE}}=1$, some post-CE separations are small enough for NSs forming from ECSNe with small kicks, to merge within a Hubble time (three simulations at $P_{\mathrm {orb}} \sim 560$~days in the left panel). 
In our models, we find that variations to the parameters of CE lead to a change in the merger fraction of almost two orders of magnitude. 

The fourth row in Fig.~\ref{fig:grids} corresponds to simulations ran with subsolar metallicity $Z = 0.1 Z_{\odot}$ and $\alpha_{\mathrm{CE}} = 1$. 
In this case, we do not find any WDs.
For these models we find more merging BNSs compared to simulations at solar metallicity. 
This is likely in part because subsolar metallicity stars are more compact. 
With a smaller donor radius, the boundary between systems that experience CE evolution and those that do not, occur at smaller initial periods than at solar metallicity.
As a result, systems that successfully eject the envelope in our simulations result in smaller post-CE orbital periods and lead to more merging BNSs.
While we find that subsolar models are more efficient at forming BNS mergers, a full astrophysical population is needed in order to properly quantify if this metallicity dependence significantly affects the predicted population of BNS mergers in the Universe.

Previous studies have shown that post-CE MT can further harden the binary system, increasing the chance of a BNS merger \citep[e.g.,][]{Tauris2015,Tauris2017, Kruckow2018}.
We find post-CE MT in our simulations {\highlight at both metallicities}.
Two examples of MT-induced orbital shrinkage are:   
(i) For the binary with the highest merger fraction at solar metallicity and $\alpha_\mathrm{CE} = 1$, we find a post-CE $2.65 M_{\odot}$ donor in a $3$~day orbital period decrease by $0.4$~days. 
This is comparable to the loss in orbital period of $0.54$~days for a binary with a $2.6 M_{\odot}$ donor in a $2.0$~day initial orbital period in \citet[][Table 1]{Tauris2015}.
(ii) For the simulation with highest merger fraction at subsolar metallicity and an initially $8 M_{\odot}$ donor, we find a $3M_{\odot}$ post-CE donor in an orbital period of roughly $0.14$~days decreases by $0.047$~days.
This is again comparable to the loss in orbital period of $0.037$~days for the binary with a $3 M_{\odot}$ donor in a $0.1$~day orbital period in \citet{Tauris2015}. 
Quantitative differences in the orbital shrinkage between \citet{Tauris2015} and our work are expected since model assumptions are different, such as the stellar winds and companion mass.   
We find that post-CE stable MT does impact the binary orbit {\highlight at both metallicities}; however, a detailed analysis of orbital period evolution for all our merging BNS is necessary to quantify the full effect of post-CE stable MT on the merger fractions we calculate.

Our grids of simulations with a NS of $1.4M_{\odot}$ do not form any BNS mergers following a phase of stable MT only.
The mass ratio of these binaries is too unequal for stable MT; the smallest mass ratio being $q=0.175$ with a donor mass of $8 M_{\odot}$ and only becoming more unequal with increasing donor mass.

\subsection{BNS simulations of binaries with $2.0M_{\odot}$ NS companions} \label{sec:2p0NS}

\begin{figure}
\centering
\includegraphics[width=0.475
\textwidth]{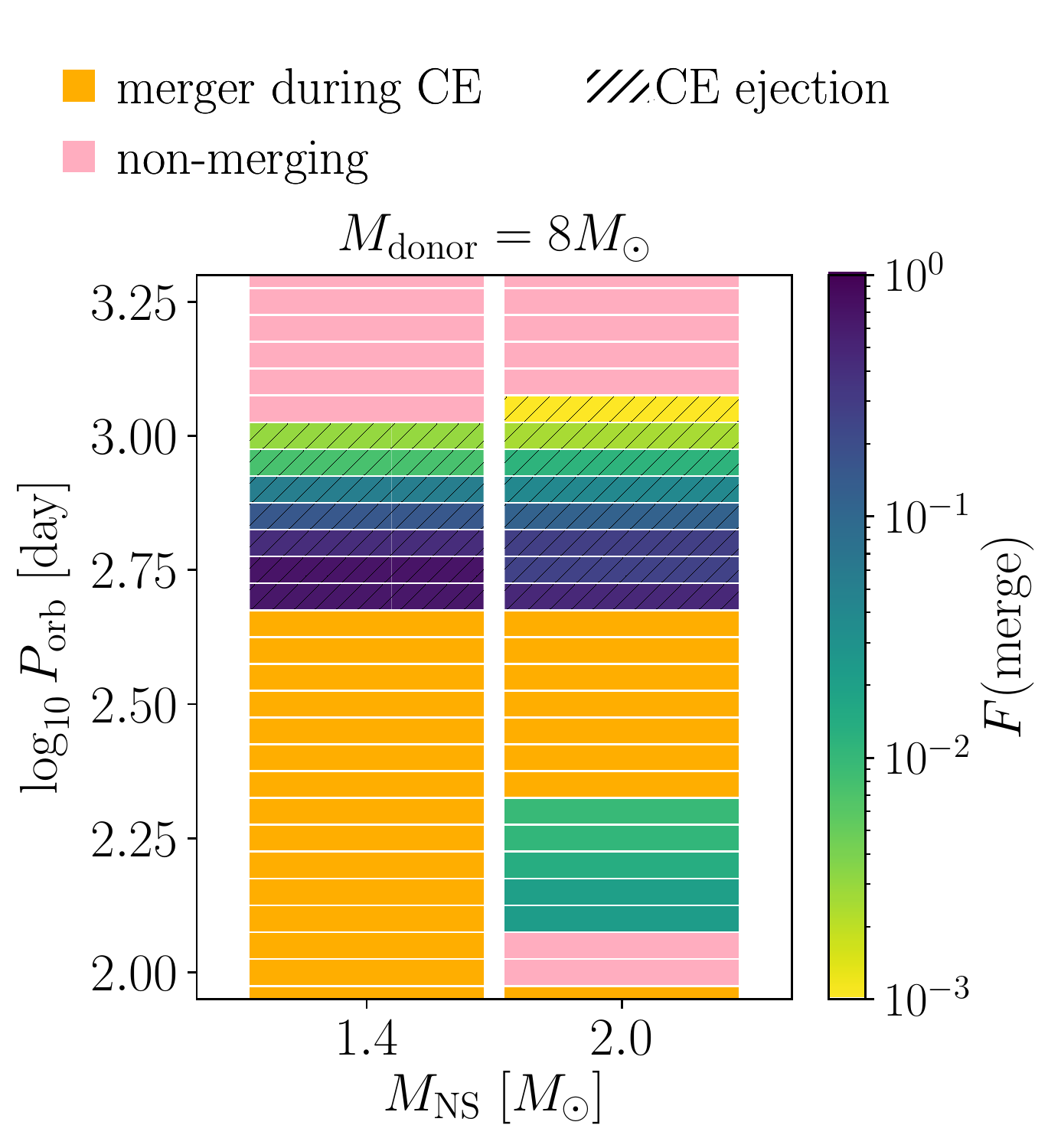}
\caption{Final outcomes for simulations at a fixed donor mass of $M_{\mathrm{donor}}=8M_{\odot}$ as a function of NS companion mass and initial orbital period at subsolar metalllicity $0.1Z_{\odot}$ and $\alpha_{\mathrm{CE}}=1$.
We show only the prescription following \cite{Tauris2015} to determine ECSN.
With a NS mass of $M_{\mathrm{NS}}=2.0M_{\odot}$, we find a non-zero BNS merger fraction following only stable MT. 
}
\label{fig:grids_heavier_NS}
\end{figure}

NSs more massive than the conventional mass of $1.4M_{\odot}$ are known \citep[e.g.,][]{Antoniadis2013,Alsing2018,Cromartie2020,Abbott2020_190425}.
Hence, we additionally consider the effects of a more equal mass ratio by varying the NS mass to $2.0M_{\odot}$ for a donor mass of $8.0M_{\odot}$, $q=0.25$.
In Fig.~\ref{fig:grids_heavier_NS} we show the outcomes of these simulations compared to the same initial conditions with a NS mass of $1.4 M_{\odot}$.
We show only the results at subsolar metalllicity $0.1Z_{\odot}$, $\alpha_{\mathrm{CE}}=1$, and following the prescription of \citet{Tauris2015} to determine ECSNe.
The simulations with a NS mass of $2.0M_{\odot}$ share the same three classes as in Fig.~\ref{fig:grids}: non-merging systems at the highest orbital periods, merging BNSs following CE around $P_{\mathrm{orb}}\sim 500$--$1000$~days, and merging during CE at lower initial orbital periods.
However, in this case, the larger NS mass now allows for stable MT at $P_{\mathrm{orb}}\sim 100$--$300$~days that leads to BNS mergers.

For the ECSN prescription following \cite{Podsiadlowski2005} all stable MT BNS receive low kicks preserving the wide orbit and do not lead to BNS mergers. 
For donors $>~8M_{\odot}$ and a NS mass of $2.0 M_{\odot}$, we do not find stable MT as in Fig.~\ref{fig:grids_heavier_NS}. 
We find that the presence of BNS mergers following stable MT only is sensitive to the initial conditions of the binary and assumptions on SN kicks. 

We focus on the case when BNS mergers do form following stable MT. 
In Fig.~\ref{fig:grids_heavier_NS}, for both NS mass and initial orbital periods $P_{\mathrm{orb}}\lesssim 100$ days, MT leads to a merger during CE.
However, with increasing initial orbital period the donor is more evolved at the onset of CE, allowing simulations with a more equal mass ratios to proceed with stable MT instead of CE. 
Similar MT stability with stellar age has been shown in \citet{ge_adiabatic_2015}. 
The more equal mass ratio not only allows for stable MT but can shrink the separation of the binary to a $\sim 7$~day orbit through loss of orbital angular momentum as a result of non-conservative MT \citep[e.g.,][]{vandenHeuvel2017}. 
However, at this binary separation a strong SN kick is still required in order for the binary to merge within a Hubble time.
At larger initial orbital periods of $P_{\mathrm{orb}}\gtrsim 200$~days, MT can occur when the donor has developed a convective envelope, which can then lead to a CE and a BNS merger at larger orbital periods $P_{\mathrm{orb}}\gtrsim 500$~days.
The merger fraction of BNS mergers following stable MT only is small, $\sim 0.01$, compared to BNS mergers following CE, which can reach $\sim 1$ for the same NS mass.
Additionally, donors that are stripped the most by stable MT (at initial orbital periods of $P_{\mathrm{orb}}\sim 100$~days) fall within the ECSN threshold and do not merge within a Hubble time. 
We therefore do not expect BNS mergers following only stable MT to be common. 

The stability of MT found here can be compared to Figure~4 in \citet{Misra2020}, where they compute a grid binary systems with donors up to $8M_{\odot}$. 
They find that for both a NS accretor of $1.3M_{\odot}$ and $2.0M_{\odot}$, MT is unstable. 
There are key differences between our simulations: \citet{Misra2020} run simulations at solar metallicity and the initial orbital periods of the binaries are $<100$~days. 
In our simulations, stable MT is found at subsolar metallicities and for larger initial orbital periods.
Most importantly, the method of \citet{Marchant2021} we use in our simulations allows for overflow out of the outer Lagrangian points without assuming it is dynamically unstable as assumed in \cite{Misra2020}.
This difference would change the systems classified as stable MT in our simulations that had overflow out of the outer Lagrangian points to CE simulations.
The trends in the outcomes for a fixed mass ratio with increasing orbital period can also be compared to the trends for BBH mergers show in Figure~8 in \citet{Marchant2021} and Figure~1 in \citet{Gallegos-Garcia2021}.
Both of these simulations show roughly the same transitions from unstable MT to a compact object binary merger following stable MT to unstable MT again with increasing initial orbital period.
These simulations show that the behavior that led to BBH mergers following stable MT only can also form BNS mergers given additional (rare) conditions like a large SN kick.

The mass ratio of the binary, combined with the initial orbital period, primarily determines the evolutionary fate of the binary in our grids. 
Our selected range in mass ratio and orbital period show a clear dominance of CE as the favored evolutionary channel with a switch to stable MT under rare conditions. 
A less massive NS will extend our range to more unequal initial mass ratios, where CE will be unavoidable.
On the more massive end, there is currently no unambiguous evidence of NSs with mass $\gtrsim 2.2 M_{\odot}$ \citep{Alsing2018,Shao2020_NS_cutoff,Most2020,Abbott2021_pop_through_GWTC3,Zhu2023}.
Therefore, we do not expect that considering a realistic population of NS masses will affect our finding that CE is the dominant evolutionary channel for BNS mergers.

\section{Conclusions}\label{sec:conclusions}

We have used grids of \MESA{} simulations with detailed methods for MT and CE evolution to study the formation of BNS mergers.
Our simulations follow the evolution of a donor star within a mass range of $7$--$20M_{\odot}$ and a $1.4M_{\odot}$ and $2.0M_{\odot}$ NS at varying orbital periods. 
In our models we varied metallicity, the CE efficiency parameter $\alpha_{\mathrm{CE}}$, and the SN prescription for determining which progenitors explode as ECSNe.
For each BNS simulation, we sampled kick velocities from a Maxwellian distribution with a velocity dispersion determined by the type of SN. 
To determine the efficiency of forming BNS mergers we calculated a \mergprob{}, the fraction of kicks sampled that led to BNS mergers within a Hubble time per simulation.  

Unlike the case of BBH mergers, which form predominantly via stable MT \citep{Marchant2021,Gallegos-Garcia2021}, we find that BNS systems form predominantly following CE \citep{Postnov2014}.
Our models with solar metallicity, {\highlight tend to have wide post-CE separations.} 
We find that larger kicks, although they can disrupt the binary, {\highlight tend to produce more} BNS mergers since they can impart large eccentricity on our wide post-CE separations and merge within a Hubble time; without strong kicks our post-CE orbital periods {\highlight tend to be} too wide for the system to merge within a Hubble time. 
{\highlight At lower $\alpha_\mathrm{CE}$, we begin to find post-CE separations small enough to merge without large natal kicks.}
When varying the metallicity of the donor star to subsolar metallicity, $Z=0.1Z_{\odot}$, we find higher fractions of BNS mergers. 
This is likely because the more compact, lower-metallicity stars have successful CE phases (ejecting the envelope) at shorter orbital periods than solar metallicity stars. 
We find that the post-CE separations of these binaries are smaller compared to our solar metallicity models and thus are more likely to lead to BNS mergers.

Our simulations may show different trends with assumed stellar and binary physics parameters compared to previous studies.  
The dependencies on metallicity we find are in contrast to \citet{GiacobboMap2019_ecsn} and \citet{Giacobbo2018_impact_of_Z}, which show that the efficiency of BNS mergers either reaches a minimum near $Z=0.1 Z_{\mathrm{\odot}}$ or tends to be insensitive to metallicity, depending on CE and kick assumptions \citep{Klencki2018,Chruslinska2018,neijssel_effect_2019,Santoliquido2021,Broekgaarden2022,Iorio2022}. 
Unlike these studies, we do not preform a full population study and thus do not model the evolution prior to the formation of the first NS. 
Metallicity trends in this initial stage of evolution could impact the underlying distribution of NS--H-rich progenitors.
Nonetheless, our simulations may indicate that the formation efficiency of BNS mergers may be more sensitive to metallicity than previously thought. 
Follow-up studies using, or informed by, detailed binary simulations such as these are required to fully address this metallicity dependence. 

We find that the outcome of our simulations can be sensitive to the mass ratio of the binary systems.
When considering a more equal mass ratio we find BNS mergers following a phase of stable MT without CE but only if the newly-formed NS receives a strong SN kick. 
BNS mergers from stable MT comprise a small subset of the total BNS mergers in our grid and likely do not play a large role in the formation of BNS mergers within a Hubble time. 

The outcomes of binary evolution modeling are known to be highly sensitive to natal kick velocities.
For example, using the rapid population synthesis code \texttt{COMPAS} \citep{Stevenson2017,Riley2022}, \citet{Vigna2018} show that when all SNe in their population are given high kick velocities, they fail to reproduce observed Galactic BNSs with low eccentricities and large orbital periods,  but a bimodal kick distribution is preferred over a unimodal distribution. 
Similarly, based upon observations of Galactic BNSs, \citet{Tauris2017} suggest that while the kick velocity for the second SN explosion is favored to be small, a few systems may also be consistent with high velocity kicks. Similar results were found by \cite{Wong2010}.
Additionally, in contrast to our results, other studies have found that smaller kicks lead to larger merger rates for BNS. For example, using the rapid binary populations synthesis code \texttt{COMBINE}, \cite{Kruckow2018} found ECSN kicks a factor of $2$ smaller resulted in an increase in the rate of BNS mergers of roughly 3. 
Using the rapid population synthesis code \texttt{MOBSE} \citep{Giacobbo2018_eddington}, \citet{GiacobboMap2019_ecsn} find that the number of merging BNS systems is larger when kick velocities are sampled from a Maxwellian distribution with $\sigma_{\mathrm{ECSN}}=7$--$26~\ \mathrm{km \ s}^{-1}$ than with $\sigma_{\mathrm{ECSN}}=265~\ \mathrm{km \ s}^{-1}$. 
{\highlight For our set of models at two metallicities and CE efficiencies, we find that more BNS mergers within a Hubble time are produced when kicks are large. 
The kick distribution required to produce high \mergprobeq{} depends strongly upon the post-CE separations, which in turn depends upon the CE treatment.
}

The efficiency of CE evolution remains highly uncertain. 
Previous studies, implementing either single-star grid-based simulations or fitting formulae of stellar models, have shown that a larger $\alpha_\mathrm{CE}$ tends to increases the number of BNS mergers \citep[e.g.,][]{Kruckow2018,Broekgaarden2022}. 
This could be expected since a larger efficiency will result in more envelope ejections and therefore more possible binaries resulting in compact object mergers.
Our results show the opposite: although a larger $\alpha_\mathrm{CE}$ increases the number of successful envelope ejections, it also causes the once wide post-CE separations to become even wider and not merge within a Hubble time. 
For this reason we find a smaller $\alpha_\mathrm{CE}$ increases the fraction of BNS mergers.
For CE evolution, we do find, however, qualitatively similar results to \citet{HiraiMandel2022} for post-CE separations.
\citet{HiraiMandel2022} find wider post-CE separations compared to the classic CE energy formalism. 
Our work seems to suggest the same.
Therefore, there appears to be a building consensus that the standard CE energy formalism overestimates orbital shrinkage during CE.

Given our more detailed stellar modeling during CE, it is likely that the interplay between post-CE separations {\highlight at a particular $\alpha_\mathrm{CE}$} and natal kicks is different in our simulations compared to previous studies.
{\highlight As our primary goal is to explore the prevalence of BNS mergers formed via CE evolution and stable MT, we do not constain uncertain parameters like $\alpha_\mathrm{CE}$ or SN kick velocity distributions.
These uncertain parameters are only used to explore trends in our models.}
Both CE and SN kicks play critical roles in the merger rate predictions of BNSs and NSBHs  \citep{Giacobbo2018_impact_of_Z,Broekgaarden2021_impact_of_on_NSBH}. {\highlight They also impact orbital properties, which can be compared to Galactic observations \citep[e.g.][]{Wong2010, Tauris2017, Vigna2018}, where a lack of eccentricity may support SNe with smaller kicks.} Hence, these physical parameters must be explored in more detail in order to fully understand their impact on the formation of mergers involving NSs. 

Our simulations do not capture the evolution of the binary prior to the formation of the first NS. 
MT during the MS--MS phase has been shown to affect the internal rotation profile and envelope structure of the accreting star \citep[e.g.,][]{Renzo2021}. 
A different envelope structure in particular can affect the outcome of CE in our models. 
Capturing the effects of this MT and its longevity should be considered in future models.

Current predictions for compact object merger properties, such as the local merger rate, have large uncertainties and scatter between estimates \citep{MandelBroekgaarden2022}.
Especially in the case of mergers involving NSs, the CE phase remains unavoidable and predictions are hindered by unresolved uncertainties and modeling. 
While progress is being made in hydrodynamical simulations \citep{LawSmith2020,DeSoumi2020,Lau2022,Moreno2022,Lau2022_massive_stars}, it is still important to model the formation of BNS mergers in lower dimensions so that it is remains computationally feasible to create large populations.
It is only with large astrophysical populations that we will be able to constrain uncertainties related to binary physics in order correctly interpret and form robust predictions for future observations.

\begin{acknowledgements}

The authors thank the referee for comments on the manuscript; Meng Sun, Jeff Andrews, and Aaron Dotter for their feedback and assistance with our \MESA{} simulations; Chase Kimball for discussions on SN prescriptions; Michael Zevin and the \texttt{POSYDON} collaboration (\href{https://posydon.org}{posydon.org}) for useful discussions on stellar evolution, and Zoheyr Doctor for feedback on the manuscript. 
M.G.-G.\ is grateful for the support from the Ford Foundation Predoctoral Fellowship. 
C.P.L.B.\ acknowledges past support from the CIERA Board of Visitors Research Professorship and current support from the University of Glasgow. 
V.K.\ is supported by a CIFAR G+EU Senior Fellowship, by the Gordon and Betty Moore Foundation through grant GBMF8477, and by Northwestern University.
This work utilized the computing resources at CIERA provided by the Quest high performance computing facility at Northwestern University, which is jointly supported by the Office of the Provost, the Office for Research, and Northwestern University Information Technology, and used computing resources at CIERA funded by NSF PHY-1726951.
Input files and data products are available for download from Zenodo.\footnote{\href{https://doi.org/10.5281/zenodo.7304993}{doi.org/10.5281/zenodo.7304993}} 
\end{acknowledgements}

\software{
\texttt{MESA} \citep{Paxton2011,Paxton2013,Paxton2015, Paxton2019}; 
\texttt{MESASDK} version 20190503 \citep{townsend2019}
\texttt{Matplotlib} \citep{Hunter2007}; 
\texttt{NumPy} \citep{vanderwalt2011};
\texttt{Pandas} \citep{mckinney-proc-scipy-2010}.
}

\bibliography{ms}
\bibliographystyle{aasjournal}

\end{document}